\documentclass[aps,prl,twocolumn,superscriptaddress,showpacs]{revtex4}

\usepackage[dvips]{graphicx}
\usepackage{amsmath}
\usepackage{amssymb}
\usepackage{bm}

\newcommand{\nonum}{\nonumber}

\begin{document}

\title{The Complexity of Vector Spin Glasses}

\author{J.\ Yeo}
\affiliation{Department of Physics and Astronomy, University of Manchester,
Manchester M13 9PL, U.\ K.}
\affiliation{Department of Physics, Konkuk University, Seoul 143-701, Korea}
\author{M.\ A.\ Moore}
\affiliation{Department of Physics and Astronomy, University of Manchester,
Manchester M13 9PL, U.\ K.}

\date{\today}

\begin{abstract}
We study the annealed complexity of the $m$-vector spin glasses
in the Sherrington-Kirkpatrick limit.  
The eigenvalue spectrum of the Hessian matrix
of the Thouless-Anderson-Palmer (TAP) free energy is found to consist of a continuous 
band of positive eigenvalues in addition to an isolated eigenvalue and $(m-1)$
 null eigenvalues due to rotational invariance.
Rather surprisingly, the band does not extend to zero at any finite temperature.
The isolated eigenvalue becomes zero in the thermodynamic limit,
as in the Ising case ($m=1$), indicating that the same supersymmetry
breaking recently found in Ising spin glasses occurs in vector spin glasses. 
\end{abstract}



\maketitle

The recent renewed interest in the complexity of spin glasses has produced 
new insights into their properties \cite{ABM, CGP, susy6}. The 1980 calculation of the 
complexity of the Ising Sherrington-Kirkpatrick (SK) model by Bray and Moore (BM)
 \cite{BM80} has
 been confirmed but new features of it have now emerged.   
For that model the complexity is the average number of solutions,
$\langle N_S \rangle_J$, of the Thouless-Anderson-Palmer (TAP) equations \cite{TAP}.
The Hessian matrix associated with these solutions
(the matrix $\partial^2 F/\partial m_i\partial m_j$), where $F$ is their free energy
and $m_i=\langle S_i\rangle$ is the local magnetization has to have no negative
eigenvalues for a solution of the TAP equations to be a minimum of the TAP free energy.
The eigenvalues of this matrix have been found to form a band such that the
smallest eigenvalue in the band is
positive, together with an isolated eigenvalue which is zero in the large $N$ limit,
where $N$ is the number of sites. For $N$ large, the solutions of the TAP equations 
occur in pairs, with one solution corresponding to a minimum of the free energy and
 a small positive value for the isolated eigenvalue and the other solution 
corresponding to an (index-one) saddle point of the free energy $F$, with the
 isolated eigenvalue being small but negative. The appearance of a null eigenvalue in the 
large $N$ limit can be attributed to the breaking of a supersymmetry 
 \cite{susy6,susy1,susy2,susy3,susy4} 
-- the BRST supersymmetry \cite{BRST} -- possessed 
by the action that appears in the evaluation of $\langle N_S\rangle_J$ \cite{PR}.

To date only the complexity of the spherical $p$-spin model has been studied in as much
 detail as that of the
 Ising SK model.  For this model the
 supersymmetry is unbroken and so the isolated eigenvalue is no longer a null eigenvalue 
\cite{CLR}. The minima are genuine mimima of the free energy and not just turning points
 as in the Ising case.
 The dynamical consequences of this difference in the value of the isolated eigenvalue 
are striking. For
the Ising model, the TAP states, being turning points only of the free energy, do not 
trap the system near them for any substantial period of time, whereas in the spherical
 $p$-spin model the system evolves from its initial state and becomes trapped for ever
 near the TAP solution accessible from the initial state \cite{CK}.

In this paper, we examine the complexity of models which are often closer to
experimental spin glasses than the Ising spin glass, namely,
the $m$-vector spin glass models. They include 
the XY ($m=2$) and Heisenberg ($m=3$) spin glasses.
We studied also the eigenvalue
spectrum of the Hessian matrix of the TAP free energy for this model. 
This was previously obtained  at zero temperature in Refs.~\cite{BM81,BM82}.
We show below that the eigenvalue spectrum at {\em finite} temperature
consists of a continuous band of positive eigenvalues, together with $(m-1)$ null
eigenvalues due to rotational invariance and a further null eigenvalue (in the 
thermodynamic limit) whose origin is as for the Ising spin glass \cite{ABM}.
We had expected that the band edge would be at zero due to the rotational invariance
in these models, as found at zero temperature, but in fact there is a small but
finite gap when $0<T<T_c$.  

The $m$-vector SK spin glass model has Hamiltonian 
\begin{equation}
{\mathcal H}=-\frac{m}{2}\sum_{ij} J_{ij}{\bf S}_i\cdot{\bf S}_j,
\label{Hamiltonian}
\end{equation}
where the $m$-component 
spins ${\bf S}_i=\{S_i^\alpha\}$, $(\alpha=1,\ldots,m,~i=1,\ldots,N)$
of unit length $S_i=1$, and the interactions
$J_{ij}$ have a Gaussian distribution 
with zero mean and variance $1/N$ so that $T_c=1$ for all $m$.
 The TAP equations for the local fields ${\bf h}_i$
at temperature $T=\beta^{-1}$ are given in terms of the local magnetization 
variables ${\bf m}_i=\langle {\bf S}_i\rangle$ by \cite{BM81}
\begin{equation}
 h_i^\alpha=\sum_j J_{ij}m_j^\alpha-\beta(1-q)m_i^\alpha,
\label{tap1}
\end{equation}
where $q=N^{-1}\sum_i {\bf m}_i^2$, and 
${\bf m}_i$ is related to ${\bf h}_i$ by
$m_i^\alpha =\widehat{h}_i^\alpha L(m\beta h_i)$.
The function $L(x)=I_{\frac m2} (x) / I_{\frac{m-2}2} (x)$ 
and $I_\nu (x)$ is a modified Bessel function. Here
$\widehat {\bf h}_i ={\bf h}_i/h_i$.
The TAP equations are obtained by extremizing 
the free energy
\begin{eqnarray}
F & = & -\frac m 2 \sum_{i,j}J_{ij}{\bf m}_i\cdot{\bf m}_j
-\frac m 4 N\beta(1-q)^2 \nonum \\
&&~~~~~~~- T\frac\partial{\partial T}\left[
T\sum_i \ln z(m\beta h_i )\right]
\label{ftap}
\end{eqnarray}
with respect to $m_i^\alpha$, where $
z(x)=2\pi^{\frac m2}I_{\frac{m-2}2}(x)/(\frac x2)^{\frac{m-2}2}$ \cite{BM82}.
Using the TAP equations, (\ref{tap1}), the free energy can be written as 
$\beta F/N=N^{-1}\sum_{i}f(m\beta h_i)$ where
\begin{equation}
f(x)=-\frac 1 4 m\beta^2(1-q^2)+\frac x2 L(x)-\ln z(x).
\end{equation}

Following \cite{BM80,BM81},
the number $N_S(f)$ of solutions to the TAP equations
with free energy $f$ per spin is given by
\begin{eqnarray}
N_S(f)&=&N^2\int dq\int\prod_{i,\alpha} dm_i^\alpha
\int\prod_{i,\alpha} dh_i^\alpha |\det{\bf K}| \label{ns}\\
&&\times\prod_{i,\alpha}
 \delta\Big( h_i^\alpha-\sum_j J_{ij}m_j^\alpha+
\beta(1-q)m_i^\alpha\Big)\delta(H_i^{\alpha})
  \nonum\\
&&\times\delta\Big(Nq-\sum_i{\bf m}_i^2\Big)\,
\delta\Big(Nf-\sum_i f(m\beta h_i)\Big)  \nonum .
\end{eqnarray}
Here $H_i^{\alpha}= m_i^\alpha -\widehat{h}_i^\alpha L(m\beta h_i)$ and
 the matrix ${\bf K}$ is such that 
$K^{\alpha\beta}_{ij}=\partial H_i^\alpha/\partial m_j^\beta=
\sum_\gamma C_i^{\alpha\gamma}A_{ij}^{\gamma\beta}$, 
where
\begin{equation}
C_i^{\alpha\beta}= P_i^{\alpha\beta}\left(\frac{L(m\beta h_i)}{\beta h_i}
\right)+\widehat{h}_i^\alpha \widehat{h}_i^\beta mL^\prime(m\beta h_i) 
\label{C}
\end{equation}
with $P_i^{\alpha\beta}=\delta^{\alpha\beta}-
\widehat{h}_i^\alpha \widehat{h}_i^\beta$, and
\begin{eqnarray}
A_{ij}^{\alpha\beta}
&=&
\delta_{ij}(C_i^{-1})^{\alpha\beta}-\delta^{\alpha
\beta}\left[\beta J_{ij} - \beta^2(1-q)\delta_{ij}\right] 
\nonum\\
&& 
-\frac{2\beta^2}{N}m_i^\alpha m_j^\beta.  \label{hessian}
\end{eqnarray}
The matrix ${\bf A}$ is just the Hessian matrix of the 
TAP free energy given by $A_{ij}^{\alpha\beta}=
\partial^2 (\beta F/m)/\partial m_i^\alpha \partial m_j^\beta$.

Averaging the number of solutions (\ref{ns}) 
over the $J_{ij}$ distribution, we obtain \cite{BM81}
\begin{equation}
\frac 1N\ln\langle N_s (f) \rangle_J =-m\Lambda q -uf -\frac m{2\beta^2}
\Delta^2-m(1-q)\Delta+\ln I,
\label{nsj}
\end{equation}
where 
\begin{eqnarray}
I&=&\frac 2{\Gamma(\frac m2)}\left(\frac m{2q} \right)^{\frac m2}
\int_0^\infty dh\,h^{m-1}\exp\Big[m\Lambda L^2(m\beta h) \nonum \\
&&\quad +uf(m\beta h)-\frac m{2\beta^2 q}
\Big(\beta h -\Delta L(m\beta h)\Big)^2\Big].
\label{I}
\end{eqnarray}
The parameters $q,\Lambda,\Delta$ and $u$ were originally introduced 
as integration variables and are fixed according to the steepest descent method
to make the right hand side of Eq.~(\ref{nsj})  
stationary. The corresponding saddle point equations for these variables
can be derived in a straightforward way. We have solved them
numerically for temperatures $T<T_c$.
The calculation for the total number of solutions $\langle N_S \rangle_J
\sim e^{N\Sigma(T)}$ is done by setting $u=0$. We have evaluated $\Sigma(T)$ as 
a function of temperature $T$ for the XY and Heisenberg cases. 
We find that $\Sigma(T)$
increases with decreasing $T$, approaching the known $T=0$ values,
$\Sigma(0)=0.02328$ for $m=2$ and $\Sigma(0)=0.00839$ for $m=3$ \cite{BM81} and
vanishes as $T$ approaches $T_c$.
 
We now turn to the determination of the eigenvalues of the Hessian matrix
${\bf A}$.
In Eq.~(\ref{C}), both $P_i^{\alpha\beta}$ and $\widehat{h}_i^\alpha 
\widehat{h}_i^\beta$ are projection operators, therefore, the eigenvalue
spectrum of $C_i^{\alpha\beta}$ is simple: one eigenvalue
$mL^\prime(m\beta h_i)$ and $(m-1)$ eigenvalues 
$L(m\beta h_i)/\beta h_i$. They are both positive. 
For convenience, we denote by $B_{ij}^{\alpha\beta}$ 
the part of the matrix ${\bf A}$ without the final $O(1/N)$ term in (\ref{hessian}):
$A_{ij}^{\alpha\beta}=B_{ij}^{\alpha\beta}-\frac{2\beta^2}{N}
m_i^\alpha m_j^\beta$. The $O(1/N)$ term 
has the form of a projection operator and
does not contribute to the extensive part of the complexity $\Sigma(T)$. 
For the computation of $\Sigma(T)$, we can drop the 
the modulus sign around ${\bf K}$ in (\ref{ns}) \cite{ABM}.  
We will find below that, as in the Ising spin glass case, the projector term
produces an isolated eigenvalue of value zero \cite{ABM}.

\begin{figure}
\includegraphics[width=0.45\textwidth]{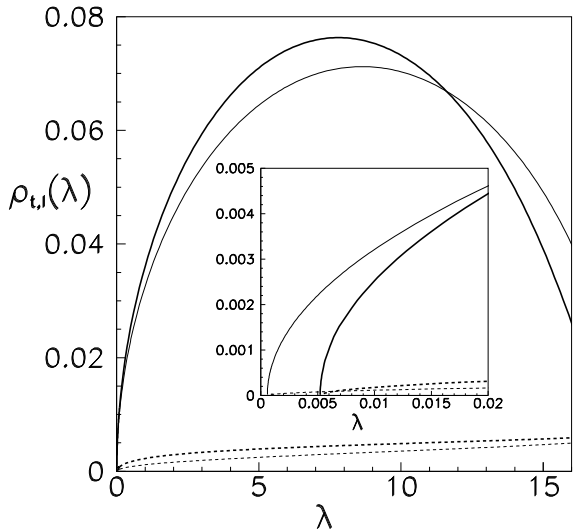}%
\caption{The transverse and longitudinal eigenvalue densities, 
$\rho_t(\lambda)$ (solid lines) and $\rho_l(\lambda)$ (dashed lines),
respectively, of the Hessian matrix for $T=0.2$ and $u=0$ (excluding the delta
functions associated with null eigenvalues). 
Thick lines correspond to $m=2$ (XY spins), while thin lines are 
for $m=3$ (Heisenberg spins). 
The inset shows the behavior
of these quantities at small values of $\lambda$, which clearly indicate
the presence of finite gaps in the spectrum.\label{fig_rho}}
\end{figure}

We calculate the eigenvalue density $\rho(\lambda)$
of the Hessian matrix ${\bf B}$ without the projector term
using the resolvent matrix
\begin{equation}
{\bf G}(\lambda)=(\lambda{\bf I}-{\bf B})^{-1},
\end{equation}
where ${\bf I}$ is the $(mN\times mN)$ unit matrix.
We can write 
\begin{equation}
\rho(\lambda)=\frac 1{N\pi} \; {\rm Im}\; {\rm Tr}\; {\bf G}(\lambda-i\delta),
\end{equation}
where $\delta$ is a positive infinitesimal.
We use the locator expansion method \cite{BM79} to calculate
$\rho(\lambda)$ in the thermodynamic limit. 
The eigenvalue spectrum consists of two branches,
transverse and longitudinal, which are denoted respectively by
$\rho_t$ and $\rho_l$. We can write
\begin{equation}
\rho(\lambda)=(m-1)\rho_t(\lambda)+\rho_l(\lambda).
\end{equation}
Each branch is 
obtained from the corresponding resolvent matrices:
$\rho_t(\lambda)=\pi^{-1}{\rm Im}\;\bar{G_t}(\lambda-i\delta)$ and
$\rho_l(\lambda)=\pi^{-1}{\rm Im}\;\bar{G_l}(\lambda-i\delta)$,
where 
\begin{equation}
\frac{1}{N}{\rm Tr}\;{\bf G}(\lambda)\equiv\bar{G}(\lambda)=
(m-1)\bar{G_t}(\lambda)+\bar{G_l}(\lambda).
\end{equation}
The functions
$\bar{G_t}(\lambda)$ and $\bar{G_l}(\lambda)$ satisfy
\begin{equation}
\bar{G_t}(\lambda)=\left\langle\left[\lambda-\beta^2(1-q)-\frac{\beta h}
{L(m\beta h)}-\beta^2\bar{G}(\lambda)\right]^{-1}\right\rangle, 
\label{gt} 
\end{equation}
\begin{equation}
\bar{G_l}(\lambda)=\left\langle\left[\lambda-\beta^2(1-q)-\frac 1
{mL^\prime(m\beta h)}-\beta^2\bar{G}(\lambda)\right]^{-1}\right\rangle ,
\label{gl}
\end{equation}
where the averages were evaluated over the weight function in the integrand 
of Eq.~({\ref{I}}). In order to solve the above equations,
we separate $\bar{G_t}(\lambda)$ and $\bar{G_l}(\lambda)$ into
real and imaginary parts. The resulting four coupled equations
are solved numerically to yield $\rho_t(\lambda)$ and $\rho_l(\lambda)$
for given temperature $T$. The results for $T=0.2$ are shown 
in Fig.~\ref{fig_rho}.
We find that, as $T$ approaches $T_c$, $\rho_t(\lambda)$ and $\rho_l(\lambda)$ become
 equal, but
in the low temperature limit $T\to 0$, however, the longitudinal eigenvalues become harder
and harder. 
This is already apparent at $T=0.2$ in Fig.~\ref{fig_rho}, where $\rho_t$
is much larger than $\rho_l$ except at large values of $\lambda$.
One of our main findings is that, for $0<T<T_c$ and
for finite $m$, the continuous part of the eigenvalue spectrum does not
extend to $\lambda=0$ as there is a very small but finite gap 
which can be seen in the inset of Fig.~\ref{fig_rho}. 
This is in contrast to the behavior at zero-temperature of the same model
\cite{BM82}, where the lower band edge is at $\lambda=0$ for all $m$,
and the large-$m$ limit where explicit solution of our equations shows that 
the spectrum is gapless at all temperatures below $T_c$ and given by 
the well-known semi-circle law:
$\rho_{m\to\infty}(\lambda)/m=\sqrt{\lambda(4\beta-\lambda)}/(2\pi\beta^2)$.

For $\lambda$ small, we can obtain analytic solutions to Eqs.~(\ref{gt})
and (\ref{gl}). Expanding these equations in powers of 
$\lambda-\beta^2[\bar G(\lambda)-\bar G(0)]$, 
we find that both longitudinal and transverse branches show a square-root
dependence but with different overall coefficients. We find
\begin{equation}
\rho_{t,l}(\lambda)\sim\frac{c_{t,l}}{\pi\sqrt{p}}\sqrt{\lambda-
\frac{x_p^2}{4p}}, \label{small}
\end{equation}
where 
\begin{eqnarray}
x_p&=&1-\beta^2\bigg[\left(1-\frac 1m\right)\langle\left(
\frac{L(m\beta h)}{\beta h}\right)^2\rangle \nonum \\
&&~~~~~~~~~~~~~~~~~ 
+\frac{1}{m}\langle\left(mL^\prime(m\beta h)\right)^2
\rangle\bigg], \\
p&=&\beta^2\bigg[\left(1-\frac 1m\right)\langle\left(
\frac{L(m\beta h)}{\beta h}\right)^3\rangle \nonum \\
&&~~~~~~~~~~~~~
+\frac{1}{m}\langle\left(mL^\prime(m\beta h)\right)^3
\rangle\bigg],
\end{eqnarray}
and
\begin{eqnarray}
&&c_t=\langle\left(\frac{L(m\beta h)}{\beta h}\right)^2\rangle
+\frac{x_p}{p}\langle\left(\frac{L(m\beta h)}{\beta h}\right)^3\rangle,
\\
&&c_l=\langle\left(mL^\prime(m\beta h)\right)^2\rangle
+\frac{x_p}{p}\langle\left(mL^\prime(m\beta h)\right)^3\rangle.
\end{eqnarray}
Note that $(m-1)c_t+c_l=m/\beta^2$ so that the total density
$\rho(\lambda)\sim m\sqrt{\lambda-x_p^2/(4p)}/(\pi\beta^2\sqrt{p})$.
Our numerical solutions for small $\lambda$ agree quite well with
the square-root behavior. In the limit $m\to 1$, 
$m_i=L(\beta h_i)=\tanh (\beta h_i)$, and therefore $x_p=1-\beta^2 \langle
(1-m_i^2)^2\rangle$ and $p=\beta^2\langle(1-m_i^2)^3\rangle$ --
the known results for Ising spin glasses \cite{plefka}.

\begin{figure}
\includegraphics[width=0.45\textwidth]{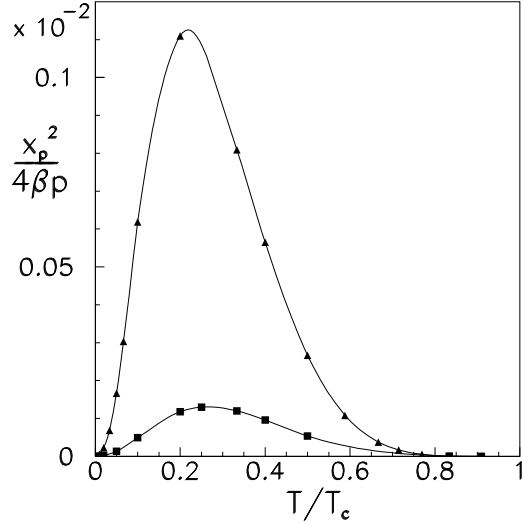}%
\caption{The parameter $x_p^2/4\beta p$ that determines
the size of the gap in the Hessian spectrum as a function 
of temperature $T$ for $m=2$ (triangles) and $m=3$ (squares). \label{fig_xp}}
\end{figure}

We have directly evaluated the parameter
which determines the gap in the expression (\ref{small}). In order to
compare with the $T=0$ result, we consider the Hessian obtained
from differentiating $F/m$ instead of $\beta F/m$. The eigenvalues will then 
 be scaled as $\lambda^\prime=\lambda/\beta$, and therefore, the gap
in the small-$\lambda^\prime$ spectrum is now determined by the gap parameter
$\frac{x_p^2}{4\beta p}$. This quantity was evaluated as a
function of temperature with the results shown in Fig.~\ref{fig_xp}.
It vanishes both at $T=0$ and $T=T_c$ as expected. The figure
also suggests that the gap becomes even
smaller when the number of components $m$ gets larger, which is consistent 
with the result of no gap at all in the large-$m$ limit. Based on the behavior
of the Ising case \cite{susy6}, we would expect $x_p$ to tend to zero for the
 TAP states whose
free energies approach those of the pure states. In other words, the TAP solutions with
the lowest free energy will be gapless at all temperatures.

At first sight, the appearance of a finite gap in the eigenvalue spectrum
at finite temperature is rather puzzling, since the continuous rotational symmetry 
of the vector spin glass model would suggest the existence of null eigenvalues associated 
with a uniform rotation of all the spins.
 One can indeed show that the Hessian matrix in the form of
Eq.~(\ref{hessian}) admits $(m-1)$ null eigenvectors in the transverse sector
 as follows: note first that
$(C_i^{-1})^{\alpha\beta}=(\delta^{\alpha\beta}-\widehat{m}_i^\alpha\widehat{m}_i^\beta)
K(m_i)/m_i+\widehat{m}_i^\alpha\widehat{m}_i^\beta K^\prime(m_i)$, where
$K(x)=L^{-1}(x)/m$. For $(m-1)$ vectors ${\bm\xi}_i$ satisfying 
$\sum_i{\bm\xi}_i\cdot\widehat{\bf m}_i=0$ and $\xi_i=m_i$, it is easy to show that
$\sum_{j,\beta}A_{ij}^{\alpha\beta}\xi_j^\beta=0$ using the defining TAP equations.

The analysis of the effect of the $O(1/N)$ projector term in the Hessian matrix 
closely follows that for the Ising case \cite{ABM}.
There is an isolated null eigenvalue outside the main band in the longitudinal
sector if
\begin{equation}
\frac{2\beta^2}{N}\sum_{i,j}\sum_{\alpha,\beta}m_i^\alpha
(B^{-1})_{ij}^{\alpha\beta}m_j^\beta \equiv2\beta^2H=1
\end{equation}
with the eigenvector $v_i^\alpha=\sum_{j,\beta}(B^{-1})_{ij}^{\alpha\beta}m_j^\beta$.
We evaluate $H$ using the expression 
\begin{eqnarray}
(B^{-1})_{ij}^{\alpha\beta}&=&\sqrt{\det {\bf B}}\int\prod_{i,\alpha}\left(
\frac{d\phi_i^\alpha}{\sqrt{2\pi}}\right)\phi_i^\alpha\phi_j^\beta \nonum \\
&&\times\exp\Big[-\frac 12 \sum_{i,j}\sum_{\alpha,\beta}
\phi_i^\alpha B_{ij}^{\alpha\beta}\phi_j^\beta\Big]
\end{eqnarray}
then insert $H$ into the integrand of (\ref{ns}). The rest of the calculation 
involves the evaluations of various Gaussian integrals and can be done
as for the Ising case \cite{ABM}, except that there are
extra indices for the vector components.
We obtain 
\begin{equation}
H=\frac{q^2 A_3}{(A_1-q)^2-A_3 (A_2-q(1-q))}, \label{H}
\end{equation}
where 
\begin{equation}
A_1=\langle L(m\beta h)\{\beta h-\Delta L(m\beta h)\}
  mL^\prime(m\beta h)  \rangle ,
\end{equation}
\begin{equation}
A_2=\langle \{\beta h-\Delta L(m\beta h)\}^2 mL^\prime
  (m\beta h) \rangle , 
\end{equation}
\begin{equation}
A_3=\langle L^2(m\beta h)mL^\prime (m\beta h) \rangle.
\end{equation} 
The averages are evaluated with respect to the integrand of (\ref{I}).
The expression for $H$ is exactly the same as the one for the Ising spin glass case
(Eq.~(10) of Ref.~\cite{ABM}) but with the quantities $A_1$, $A_2$,
and $A_3$ generalized to the $m$-vector case.
The numerical evaluation of the required integrals reveals that 
$2\beta^2 H=1$  for all $T\leq T_c$ and at all values of $u$. 
  
Therefore, in the thermodynamic limit, the Hessian matrix of the $m$-vector TAP
free energy is positive semi-definite with exactly $m$ eigenvalues equal to zero; 
the vanishing of $(m-1)$ of these eigenvalues is a consequence of rotational invariance
but the vanishing of the isolated longitudinal eigenvalue indicates that the TAP states
for vector spin glasses will behave like those for Ising spin glasses. This means that
there will be only minima and index-one saddles at large but finite values of $N$. Dynamically the 
vector spin glasses will behave more like the Ising spin glass than $p$-spin models. For 
example a state with remanent magnetization will evolve in time to a state with zero 
magnetization for all the vector glasses, whereas for $p$-spin models, remanent
magnetization remains forever -- all because  the
existence of the null longitudinal eigenvalue means that the TAP states in vector spin 
glasses are just turning points of the free energy and so are unable to trap the system
in their vicinity.





\end{document}